\documentclass[useAMS,usenatbib,usegraphicx]{mn2e}
\usepackage{mnoverride}
\newcommand{\kms}{{~\rm km\; s^{-1}}}

\newcommand{\kpc}{{~\rm kpc}}

\def \kms{~\rm{km}~{\rm s}^{-1}}
\def \kpc{~\rm{kpc}}

\usepackage[usenames,dvipsnames]{color}
\usepackage{amsmath}
\usepackage{amssymb}
\usepackage{mathptmx} 

\usepackage{setspace}

\newif\ifusehyperref
\usehyperreftrue 
\ifusehyperref
 \usepackage[unicode]{hyperref}
 \usepackage[all]{hypcap}
 \hypersetup{backref,
             bookmarks,
             dvips,
             pdftitle=Correlation of black hole and bulge masses: driven by energy but correlated with momentum,
             pdfsubject=Astrophysics \& Cosmology,
             pdfdisplaydoctitle,
             colorlinks=true,
             citecolor=Blue,
             pdfauthor=Noam Soker and Yohai Meiron,
             unicode}
 \let\OLDS\S
 \renewcommand{\S}{\textcolor{red}{\OLDS}} 
\else
 \newcommand{\href}[2]{#2}
\fi

\title{Correlation of black hole and bulge masses: driven by energy but correlated with momentum}

\author[Noam Soker and Yohai Meiron]{Noam Soker\thanks{E-mail: \href{mailto:soker@physics.technion.ac.il}{soker@physics.technion.ac.il} (NS); \href{mailto:ym@physics.technion.ac.il}{ym@physics.technion.ac.il} (YM)}
and Yohai Meiron\footnotemark[1]\\
Department of Physics, Technion -- Israel Institute of Technology, Haifa 32000, Israel}

\begin{document}

\date{Accepted 2010 October 1. Received 2010 September 18; in original form 2010 May 31}

\pagerange{\pageref{firstpage}--\pageref{lastpage}} \pubyear{2010}

\maketitle

\label{firstpage}

\begin{abstract}
We use a recent sample of 49 galaxies to show that there is a proportionality
relation between the black hole mass $M_{\rm BH}$ and the quantity $\mu\equiv
M_{\rm G}\sigma/c$, where $M_{\rm G}$ is mass of the spheroidal stellar
component and $\sigma$ is the stellar velocity dispersion. $\mu$ is called the
{\it momentum parameter} and the ratio is $M_{\rm BH} / \mu \approx 3.3$. This
result is applied to the penetrating-jet feedback model which argues that the
correlation that holds is with a momentum-like parameter, although this feedback
mechanism is based on energy balance.
\end{abstract}

\begin{keywords}
black hole physics -- galaxies : bulges
\end{keywords}

\section[Introduction]{\caps{Introduction}}
\label{sec:intro}

Relations between a supermassive black hole (SMBH) mass, $M_{\rm
BH}$, and other properties of its host galaxy have been studied
intensively in the last decade. Two of the galactic properties
most commonly correlated with $M_{\rm BH}$ are the stellar
mass of the spheroidal component (which we refer to as the bulge), $M_{\rm
G}$ (e.g. \citealt{KormendyRichstone1995};
\citealt{Magorrian_etal1998}; \citealt{Laor2001};
\citealt{Hu2009}; \citealt{Graham2009}),
and it's stellar velocity dispersion, $\sigma$
(e.g. \citealt{Gebhardt_etal2000}; \citealt{MerrittFerrarese2001};
\citealt{Graham2008a}; \citet{Graham2008b}; \citealt{Hu2008};
\citealt{Shen_etal2008}; \citealt{Gultekin_etal2009}).
These relations are often assumed to have the form of a power law (i.e.
linear when plotted on a log-log scale). However, due to the large
scatter there is still no consensus on the best parameters of the
different relations. For example, despite some claims for a
proportionality relation between the SMBH and bulge mass,
\citet{Laor2001} found that the ratio $M_{\rm BH}/M_{\rm G}$
increases with mass. This conclusion was strengthened by some
models and simulations (e.g. \citealt{ShabalaAlexander2009}).

The more significant disagreement is on which fundamental galactic property is
behind the correlations. In a recent paper, \citet{Feoli_etal2010} studied three
samples of SMBH masses and their host galaxies. They argued that $M_{\rm BH}$ is
better correlated with the energy parameter, $M_{\rm G} \sigma^2$, than with
$M_{\rm G}$ or with $\sigma$ alone. Behind this comparison stands the view that
the feedback between the accreting SMBH and its environment is driven by energy.

Results from recent years show that the process of galaxy formation requires
another energy source to that of the gravitational energy of the galaxy, not
only to heat the gas, but also to expel large quantities of it out of the galaxy
(e.g. \citealt{Bower_etal2008}). Momentum in the relativistic jets alone is not
sufficient to expel (accelerate to escape velocity, $\sim$ few $\sigma$)  a
considerable amount of mass ($\ga M_{\rm G}$). The maximum momentum that can be
released in a relativistic jet (which is the expected case for an outflow
launched by BH accretion), $\eta M_{\rm BH} c$, is generally smaller than
$M_{\rm G} \sigma$, the approximate momentum needed to expel large quantities of
gas\footnote{For a typical mass ratio of $M_{\rm G} \sim 10^3 M_{\rm BH}$, and a
velocity dispersion of $\sigma \sim 200 \kms$; $\eta \approx 0.1$ is the energy
fraction that is liberated by the accreted mass.}.

\citet{SilkNusser2010} argue that radiation momentum is also incapable of
accounting for the correlation. Let us consider \citet{King2003} as an example
of a model based on expulsion of gas by radiation momentum. In that model, an
expression for the BH mass is derived (equation 15 there) where a mass of
$M_{\rm BH} = 1.5 \times 10^{8}~{\rm M}_\odot$ is obtained for $\sigma=200 \kms$.
For this value of $\sigma$, the gas mass inside a typical radius of $10 \kpc$ is found
to be $3 \times 10^{10}~{\rm M}_\odot$ according to King's equation (4).
If this is the amount of gas expelled by radiation momentum $M_{\rm expel}$, then the
ratio of the expelled mass to the BH mass is $M_{\rm expel}/M_{\rm BH}\sim 200$. while from
This ratio from observations it is in fact $\sim 500$.
This ratio is probably even much lower in King's model as he takes the cosmological value of 0.16 for the baryon fraction, whereas this fraction is much larger (even close to unity) in the bulges of spiral galaxies. Another problem (though less severe) with this model is that one does not expect the ransfer of radiation momentum to mechanical momentum to be 100 per cent efficient.
Thus, we conclude that radiation momentum can expel only a small amount of gas, and that the correlation cannot be explained in King's model.

On the other hand, there is more than enough energy released from the accretion
process to expel the gas (by a factor of $\sim 50$), or $\eta M_{\rm BH} c^2 \gg
M_{\rm G} \sigma^2$. The problem with energy is the opposite: if deposition of
jets' energy into the ISM is too efficient, almost no star formation will occur.

Thus, a successful theory should be based on a feedback mechanism in which the
deposition of the jets' energy to the ISM is regulated; it should also be able
to derive the mathematical form of the SMBH--galaxy relation from the basic
properties of the galaxy and its active galactic nucleolus (AGN). The
penetrating-jet feedback mechanism \citep{Soker2009} has these two attributes.
In this model, the energy transfer from the jets to the ISM is efficient enough
to expel large amounts of gas and suppress star formation only when the jets are
stopped within the bulge, and do not propagate to large distances by penetrating
through the ISM. \citet{Soker2009} found that this requirement leads to a
proportionality relation between the BH mass and the {\it momentum parameter}
\begin{eqnarray}
\mu\equiv M_{\rm G}\frac{\sigma}{c}
\label{eq:mu}
\end{eqnarray}
as detailed in Section \ref{sec:model}.

Motivated by the recent data and the penetrating-jet feedback mechanism, we
carry out the present study. In Section \ref{sec:correlation} we examine the
correlation between $M_{\rm BH}$ and $\mu$. We ignore the differences between
elliptical galaxies, classical bulges and pseudo-bulges (e.g.
\citealt{GadottiKauffmann2009}; \citealt{Nowak_etal2010}; \citealt{Hu2009}). In
Section \ref{sec:feedback} we apply the results to the jet penetrating model and
derive the approximate amount of momentum carried by the jets. We summarize in
Section \ref{sec:summary}.

\section[The penetrating-jet feedback mechanism]{\caps{The penetrating-jet feedback mechanism}}
\label{sec:model}

In the penetrating-jet feedback mechanism the energy transfer from the jets is
efficient enough to expel large amounts of gas and suppress star formation only
when the jets are stopped within the bulge, and do not propagate to large
distances by penetrating into through the gas.
The condition for jet-stopping is that the time required for the jet to propagate through the surrounding gas and break out of it must be longer than the jet crossing time (the time it takes a blob of material to transverse the jet's width). This situation is analogous to an attempt to pierce a hole through a plank of wood with a drill, when the drill moves horizontally and does not stay over one spot for enough time. This is referred to as the {\it non-penetration condition}.

Mathematically, the jet crossing time at a typical radius $r_{\rm s}$ (where the cooling surrounding mass $M_{\rm s}$ resides) is
\begin{eqnarray}
\tau_{\rm c}\equiv\frac{2 r_{\rm s} \tan{\theta}}{v_{\rm rel}}\approx\frac{2 r_{\rm s} \theta}{v_{\rm rel}}
\end{eqnarray}
where $\theta$ is the half opening angle of the jet and $v_{\rm rel}$ is the relative transverse motion between the inflowing mass and the SMBH ($v_{\rm rel} \sim \sigma$). We assumes the jet to be narrow, thus $\tan\theta\approx\theta$.

The penetration time is $\tau_{\rm p}\equiv r_{\rm s}/v_{\rm h}$ where $v_{\rm h}$ is the velocity of the head of the jet, which is derived from momentum balance (ram pressure balance of the jet and the ambient gas), and is given by
\begin{eqnarray}
v_{\rm h}=\sqrt{\frac{2\dot{M}_{\rm f} v_{\rm f} \sigma}{\dot{M}_{\rm s} \theta^2}},\label{eq:v_h}
\end{eqnarray}
where the subscript `f' (for `flow') is used to denote magnitudes associated with the jets (thus $\dot{M}_{\rm f}$ is the mass outflow rate of the two jets together and $v_{\rm f}$ is its velocity). $\dot{M}_{\rm s}$ is the mass inflow rate from $\sim 1\kpc$ scale. Equation (\ref{eq:v_h}) was derived under the assumptions of: supersonic motion, $v_{\rm f}\gg v_{\rm h}$, and the stronger assumption that the inflowing speed of $\dot{M}_{\rm s}$ is $\sigma$.

The non-penetration condition $\tau_{\rm c}\lesssim\tau_{\rm p}$ leads to the following inequality
\begin{eqnarray}
\dot{M}_{\rm s} \gtrsim \frac{\sigma c}{v_{\rm rel}^2} 8 \eta_p \dot{M}_{\rm BH}\label{eq:Ms-limit}
\end{eqnarray}
where $\eta_p$ relates the rate of mass outflow $\dot{M}_{\rm f}$ to the rate of accretion or BH growth $\dot{M}_{\rm BH}$
\begin{eqnarray}
\dot{p}_{\rm f} \equiv \dot{M}_{\rm f} v_{\rm f} \equiv \eta_p \dot{M}_{\rm BH}c.
\label{eq:eta_p}
\end{eqnarray}
$\dot{p}_{\rm f}$ is the rate of momentum carried by the two jets (that we term momentum discharge). Note that equation (\ref{eq:eta_p}) is equivalent to the familiar relation for the
total energy transfer
\begin{eqnarray}
\dot{E}_{\rm BH}=\eta \dot{M}_{\rm BH} c^2,\label{eq:eta}
\end{eqnarray}
where $\dot{E}_{\rm BH}$ is the total energy released by the BH per unit time,
and $\eta\approx0.1$ is the accretion efficiency. The total energy is released
in both radiation and mechanical energy of the jets. The fraction of kinetic
energy carried by the jets is considered in Section \ref{sec:feedback}.

Equation (\ref{eq:Ms-limit}) must in fact be an approximate equality: if the inflow rate is above the value of the right hand side, the deposition of energy by the jets is efficient enough to expel the mass back to large distances and heat it. The inflowing mass that is not expelled by the jets is assumed to form stars in the bulge. Thus, time integration of both sides of the equation (\ref{eq:Ms-limit}) (with equality sign; we also substitute $v_{\rm rel}=\sigma$), leads to the following relation (\citealt{Soker2009},
\hyperlink{Soker2010}{\textcolor{Blue}{2010}})
\begin{eqnarray}
M_{\rm BH}=\frac{1}{8 \eta_p} \mu,
\label{eq:cor1}
\end{eqnarray}
where $\mu$ is the momentum parameter (defined in equation \ref{eq:mu}) which
has units of mass.

Feedback mechanisms have been discussed by many authors in the past, both to
suppress gas cooling in cooling flow clusters (e.g. \citealt{BinneyTabor1995};
\citealt{NulsenFabian2000}; \citealt{Reynolds_etal2002};
\citealt{OmmaBinney2004}; \citealt{SokerPizzolato2005}) and to suppress star
formation during galaxy formation (e.g. \citealt{SilkRees1998};
\citealt{Fabian1999}; \citealt{King2003}; \citealt{Croton_etal2006};
\citealt{Bower_etal2008}; \citealt{ShabalaAlexander2009}; \citealt{Soker2009},
\hyperlink{Soker2010}{\textcolor{Blue}{2010}}). The penetrating-jet feedback
mechanism does not make use of the Eddington luminosity limit, while some
authors do (e.g. \citealt{SilkRees1998}; \citealt{King2003}). Most models (e.g.
\citealt{SilkRees1998}; \citealt{Fabian1999}) do not consider the geometry
explicitly, while in the penetrating-jet feedback mechanism the geometry of the
narrow jets and the motion of their source are key issues; these introduce the
factor $1/8$ in equation (\ref{eq:cor1}).

In the penetrating-jet feedback mechanism, the fast jets have two modes of
interaction with the surrounding gas: if the jets penetrate through the ISM gas,
they deposit most of their energy at large distances, and thus cannot expel gas;
if they cannot penetrate, they are shocked, and a hot bubble is formed. If the
radiative cooling time of the hot bubble is longer than the flow time (or
acceleration time), it efficiently accelerates the surrounding gas and expels
it. Namely, in order to suppress star formation it is necessary that the jets do
not penetrate through the ISM. For that, the jets should encounter new material
before they escape. This requires that there is a transverse velocity component
between the jets and the ambient gas with which it interacts. The physical
condition for efficient expulsion of the gas is that the typical time for the
jets--ISM relative motion to cross the jet's width in the transverse motion at a
radius $r_s$ (from the central source) should be shorter than the penetration
time at radius $r_s$.

\section[The correlation]{\caps{The correlation}}
\label{sec:correlation}

\subsection{Methods and Sample}
\label{subsec:methods}

The $M_{\rm BH}$--$\mu$ relation is initially assumed to be a power law of the
form
\begin{eqnarray}
\log_{10}M_{\rm BH}=\alpha + \beta\log_{10}\mu.
\label{eq:corbm}
\end{eqnarray}
We use a least squares estimator of linear relations like {\tt fitexy} of
\citet{Press_etal1992}, that takes into account measurement errors in both
coordinates. Like \citet{Tremaine_etal2002}, we add a constant residual error
$\epsilon_0$ ({\it intrinsic scatter}) until $\chi_r^2=1$ is obtained (see
appendix). However, while \citet{Tremaine_etal2002} have assumed the intrinsic
scatter to be on the BH mass alone, we consider other possibilities as detailed
below. \citet{Novak_etal2006} found based on Monte Carlo simulations that this
method estimates the slope with the least bias and variance. The maximum
likelihood estimator used by \citet{Gultekin_etal2009} gives similar results to
the above, but there is no freedom to manually set $\epsilon_0$ and study how it
changes the results. The ability to vary $\epsilon_0$ (or alternatively,
$\chi_r^2$) is important because the result of the fit is sensitive to the
stated measurement errors. First we use this method to show that the slope
$\beta$ of the relation in study is reasonably close to 1, and later we force
$\beta=1$ and find the intercept $\alpha$.

We use three models for the intrinsic scatter: in the $y$-scatter model we
assume (like \citealt{Tremaine_etal2002}) that $\epsilon_0$ is the residual
variance in $\log_{10}M_{\rm BH}$ (which is traditionally the $y$ coordinate);
in the $x$-scatter model, $\epsilon_0$ is the scatter in the galactic property
in question, in our case $\log_{10}\mu$; in the orthogonal scatter model,
residual errors are added to both coordinates, such that the combined error is
$\epsilon_0$ in the direction perpendicular to the ridge line of the relation.
The estimators of the three models are given in the appendix. The actual value
of $\epsilon_0$ is meaningful only within the three models, and it is generally
wrong to compare the values obtain from each one. We also note that the
assumption that the scatter (in whatever direction) is constant throughout the
relation is made out of ignorance and may not represent the real situation.

It is impossible to avoid specifying the direction of the intrinsic scatter
\citep{Novak_etal2006}, and an extremely biased result is obtained if the
calculation is performed under the assumptions that the residual variance is in
the wrong coordinate. It is also claimed by \citet{Novak_etal2006} that the
direction of the scatter depends on whether it was the BH which affected the
host galaxy property or vice versa. In the penetrating-jet feedback mechanism,
the former is true (because the jets launched by the BH affect the gas in the
galaxy), and thus the $x$-scatter model better suits the theory. However,
acknowledging that both $M_{\rm BH}$ and the momentum parameter are affected in
a non trivial manner by phenomena such as mergers, we do not give the $x$-scatter
model any preference, and regard the orthogonal scatter model as a compromise
between the other two.

For simplicity we use the S sample of \citet{Gultekin_etal2009}, which includes
49 measured BH masses but no upper limits. In that paper, two different mass
measurements are stated for both NGC1399 and NGC5128; we take the (geometric)
average mass of each one, and the uncertainty ranges are combined. The $M_{\rm
G}$ values are taken from table 3 in \citet{Feoli_etal2010}.

Since our estimators cannot deal with asymmetric measurement errors, we adopt
the common practice of taking
\begin{eqnarray}
\delta y = {\textstyle\frac{1}{2}}\left[\log_{10}M_{\rm high}-\log_{10}M_{\rm
low}\right]
\end{eqnarray}
where $\delta y$ is the uncertainty in the logarithmic BH mass, and $M_{\rm
high}$ and $M_{\rm low}$ are the published limits of the 68 per cent confidence region.
The error on all $M_{\rm G}$ values is taken to be 0.18 dex; this dominates over
the error in $\sigma$ (which is typically $\sim0.02$ dex) in the uncertainty in
momentum parameter.

\subsection{Results}
\label{subsec:results}

\begin{figure}
\begin{center}
\includegraphics[width=1\columnwidth]{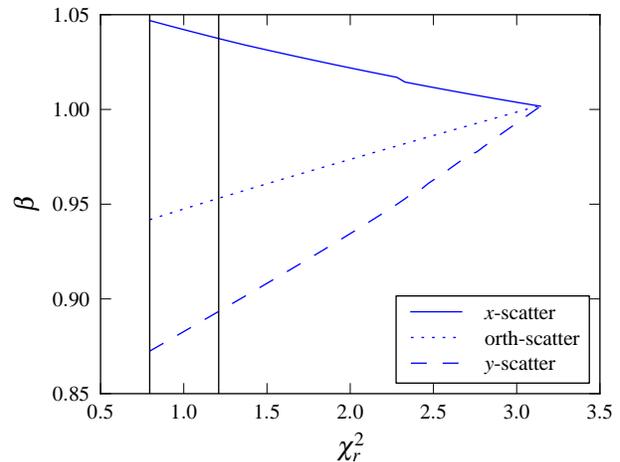}
\end{center}
\caption{The best fitting slope $\beta$ in equation (\ref{eq:corbm}) as a
function of the demanded $\chi_r^2$ in the three scatter models. The solid line
is the $x$-scatter model, the dotted line is the orthogonal scatter model, and
the dashed line is the $y$-scatter model. The two vertical lines mark the 68 per cent
confidence interval of the (reduced) $\chi^2$-distribution with 47 degrees of
freedom, centred at $\chi_r^2=1$. Note that this interval does not actually
give the uncertainty on $\beta$, which is calculated according to the algorithm
of \citet{Press_etal1992}. The best fitting slope and intercept for $\chi_r^2=1$
are given in Table \ref{tab:slopes}.}
\label{fig:slope}
\end{figure}

Fig. \ref{fig:slope} shows the best fitting slope obtained for equation
(\ref{eq:corbm}) using the three different scatter models, as a function of the
achieved $\chi_r^2$. The solid line is the $x$-scatter model (scatter in the
momentum parameter), the dotted line is the orthogonal scatter model, and the
dashed line is the $y$-scatter model (scatter in the SMBH mass). In the point on
the right where the three lines meet, $\chi^2$ is maximal and there is no
residual variance in any direction ($\epsilon_0=0$). The two vertical lines mark
the 68 per cent confidence interval of the (reduced) $\chi^2$-distribution with 47
degrees of freedom, centred at $\chi_r^2=1$. The best fitting values for
$\alpha$ and $\beta$ for $\chi_r^2=1$, and the obtained $\epsilon_0$, are given
in Table \ref{tab:slopes}. While the values of $\alpha$ vary greatly and have
large errors, the errors on all slopes are at the $\sim8$ per cent level, and all are
consistent with 1 to within two standard deviations.

\begin{table}
\begin{center}
\caption{The best fitting values for equation (\ref{eq:corbm}) in the three
models of intrinsic scatter. In all models, $\epsilon_0$ was increased until
$\chi_r^2=1$ was achieved. The values for the intercept $\alpha$ vary greatly
and have large errors, but all slopes are consistent with 1 to within two
standard deviations. Note that $\epsilon_0$ has a somewhat different meaning in
each model (see text), and does not indicate goodness-of-fit.}
\label{tab:slopes}
\begin{tabular}{lccc}
\hline\hline
Model & $\alpha\pm\delta\alpha$ & $\beta\pm\delta\beta$ & $\epsilon_0$ \\
\hline
$x$-scatter  & $0.2\pm0.6$ & $1.04\pm0.08$ & 0.32\\
$y$-scatter  & $1.4\pm0.5$ & $0.88\pm0.07$ & 0.31\\
orth-scatter & $0.9\pm0.6$ & $0.95\pm0.08$ & 0.32\\
\hline
\end{tabular}
\end{center}
\end{table}

\begin{figure}
\begin{center}
\includegraphics[width=1\columnwidth]{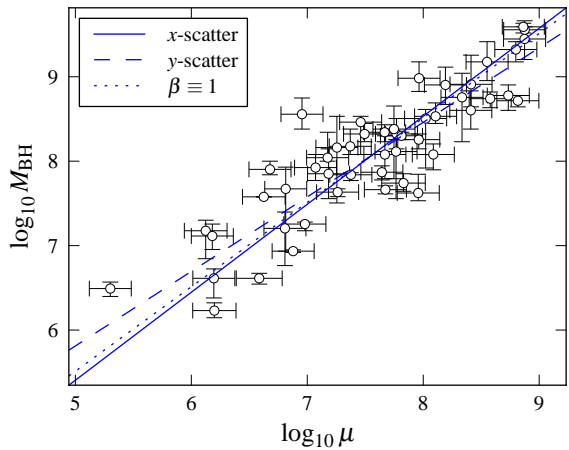}
\end{center}
\caption{The $M_{\rm BH}$--$\mu$ correlation. $\mu$ is the momentum parameter
defined in equation (\ref{eq:mu}). Both axes are logarithmic in $\rm{M}_\odot$.
The data from the \citet{Gultekin_etal2009} S sample. The solid line and the
dashed line are the best fitting straight lines according to the $x$- and
$y$-scatter models respectively. The dotted line here has a slope of $\beta=1$
and the intercept is given by equation (\ref{eq:alpha}).}
\label{fig:data}
\end{figure}

This result justifies trying to modify the assumption that the $M_{\rm
BH}$--$\mu$ relation is a power law, and instead assume that it is a linear
relation ($\beta\equiv1$). We used a similar method to {\tt fitexy} to obtain
$\alpha$, as detailed in the Appendix. The forcing of the slope to 1 renders the
three scatter models equivalent. The residual error is taken to be the average
value obtained in Table \ref{tab:slopes}. The resulting intercept is
\begin{eqnarray}
\alpha = 0.51 \pm 0.07.\label{eq:alpha}
\end{eqnarray}
The uncertainty in $\alpha$ is obtained using this method is much smaller than
before, because there is much less freedom to change $\alpha$ while minimizing
$\chi^2$. The SMBH mass is plotted as a function of the momentum parameter in
Fig. \ref{fig:data}, where the best fitting lines are also shown. The solid line
corresponds to the $x$-scatter model, the dashed line corresponds to the
$y$-scatter model, and the dotted line has a fixed slope set to $\beta=1$ and an
intercept given by equation (\ref{eq:alpha}).

The motivation to check the $M_{\rm BH}$--$\mu$ correlation comes from the
penetrating-jet feedback mechanism that gives both the slope $\beta$ and the
intercept $\alpha$ from basic parameters of the systems; one of these parameters
is determined here (see Section \ref{sec:feedback}). Other models exist, but in many
the slope and/or the intercept are not given from basic parameters of the
system. \citet{Hopkins_etal2007}, for example, take the accretion rate based on
an Eddington-limited prescription based on Bondi-Hoyle-Lyttleton accretion
theory. The Bondi-Hoyle-Lyttleton accretion theory has the disadvantage that it
cannot maintain a feedback in real time \citep{Soker2010}. Their theoretical
explanation is based on pressure-driven wind, and they expect a relation of the
form $M_{\rm BH} \propto M_{\rm G}^{1/2} \sigma^{2}$. This is not compatible
with observations, as they themselves find.

\subsection{Comparison}
\label{subsec:comparison}

We wish to compare the $M_{\rm BH}$--$\mu$ correlation we showed with
correlations with other quantities such as $\sigma$. However, by setting
$\chi_r=1$ to account for the intrinsic scatter, one makes it impossible to use
common goodness of fit tests. $\epsilon_0$ itself would be a fair indicator of
goodness of fit only if the scatter is in the same quantity, i.e. $M_{\rm BH}$,
in each relation; but we have no good indication that that is the case. Let us
nevertheless investigate other relations using the same fitting method described
in Section \ref{subsec:methods} and the $y$-scatter model (so that the scatter is in
the BH mass). We use the same dataset and assume all relations we test have the
same form as equation (\ref{eq:corbm}).

Table \ref{tab:comparison} gives the slope (with formal error) and the intrinsic
scatter for different popular relations. The slopes were calculated with the
$y$-scatter model, which we do not prefer, so that the scatter is in
$\log_{10}M_{\rm BH}$ in all tested correlations. We note that our results for
the $M_{\rm BH}$--$\sigma$ relation are very close, but not identical, to those
appearing in table 6 of \citet{Gultekin_etal2009}, where $\beta=4.06 \pm 0.37$
and $\epsilon_0=0.40$ (third row there, titled {\tt T02ind}). These numbers were
obtained using the same method and data. However, it is possible that the
discrepancy comes from different treatment of NGC1399 and NGC5128 (the galaxies
with two $M_{\rm BH}$ measurements for each); \citet{Gultekin_etal2009} take all
measurement and give each one half the weight, while in this work we average the
values and expand the uncertainty ranges.
We note that the sample of \citet{Gultekin_etal2009} is basically the same
as that of \citet{Graham2008b}, who find the slope to be 4.28, which is somewhat closer to the value we find.

All values of $\epsilon_0$ for the tested correlations are comparable; we give a
crude estimate of the error in $\epsilon_0$ to be $\approx0.05$ in all four
cases. This is determined by using the formula
\begin{eqnarray}
\delta\epsilon_0 = {\textstyle\frac{1}{2}}[\epsilon_0(\chi_{\rm
low}^2)-\epsilon_0(\chi_{\rm high}^2)],
\end{eqnarray}
where $\epsilon_0(\chi_{\rm low}^2)$ is the intrinsic scatter that produces a
$\chi^2$ at the low limit of the 68 per cent confidence range of the
$\chi^2$-distribution (shown in Fig. \ref{fig:slope}); $\epsilon_0(\chi_{\rm
high}^2)$ corresponds to the high limit of this range. The formal error in the
scatter in the $M_{\rm BH}$--$\sigma$ relation as found by
\citet{Gultekin_etal2009} using their maximum likelihood method is
$\approx0.06$.

Given that the shape of the scatter is unknown and that the assumptions made
were very simplistic, the uncertainty in $\epsilon_0$ may be much higher than
our estimate. Thus, all tested correlations seem to be equally good.

\begin{table}
\begin{center}
\caption{The slope (with error) and intrinsic scatter in correlations of the BH
mass with four parameters. All relations are assumed to be linear in the log-log
plane. The results were obtained using the $y$-scatter model, which we do not
prefer. The estimated error in $\epsilon_0$ in all cases is 0.05 (see text),
indicating that all correlations are equally good for the current data.}
\label{tab:comparison}
\begin{tabular}{lccc}
\hline\hline
Parameter & $\beta\pm\delta\beta$ & $\epsilon_0$\\
\hline
$\mu$               & $0.88 \pm 0.07$ & $0.31$\\
$\sigma$            & $4.18 \pm 0.37$ & $0.38$\\
$M_{\rm G}$         & $1.07 \pm 0.09$ & $0.33$\\
$M_{\rm G}\sigma^2$ & $0.74 \pm 0.05$ & $0.31$\\
\hline
\end{tabular}
\end{center}
\end{table}

\section[Implications for the penetrating-jet feedback mechanism]{\caps{implications for the penetrating-jet feedback mechanism}}
\label{sec:feedback}

We now use the results of Section \ref{subsec:results} to estimate the value of
$\eta_p$. We recall that $\eta_p$ is defined in equations (\ref{eq:cor1}) and
(\ref{eq:eta_p}), and is the ratio of the momentum discharge carried to $\dot
M_{\rm BH} c$. Assuming a proportionality relation between $M_{\rm BH}$ and
$\mu$, the ratio is $M_{\rm BH} / \mu = 10^{\alpha}=3.26\pm0.51$ according to
equation (\ref{eq:alpha}). Substituting this value in equation (\ref{eq:cor1})
gives
\begin{eqnarray}
\eta_p = 0.038 \pm 0.006.
\label{eq:pm1}
\end{eqnarray}
We thus derived the average value of a fundamental parameter characterizing jets
launched by an AGN during the phase of SMBH growth at galaxy formation.

From equations (\ref{eq:eta_p}) and (\ref{eq:eta}) it is possible to derive the
fraction $f$ of the total energy that is carried by the jets. Namely, the
kinetic power of the jets divided by the total power of the accreting SMBH as
given by equation (\ref{eq:eta}). Substituting the relativistic expressions for
the jets' kinetic energy and momentum, we find
\begin{eqnarray}
f = \frac{\dot{E}_{k,\rm f}}{\dot{E}_{\rm BH}}
= \frac{\eta_p}{\eta}\frac{c}{v_{\rm f}}\left( 1-\sqrt{1-\frac{v_{\rm
f}^2}{c^2}} \right)
\label{eq:f}
\end{eqnarray}
where $v_{\rm f}$ is the velocity of the jets. In the relativistic limit
($v_{\rm f}\rightarrow c$), the fraction $f$ of kinetic energy approaches a
constant ${\eta_p}/{\eta}=0.38$ (for $\eta=0.1$); in the non-relativistic limit
($v\lesssim 0.5c$) equation (\ref{eq:f}) becomes $f=\frac{\eta_p}{2\eta}
\frac{v_{\rm f}}{c}$. Namely, in the relativistic limit and during the phase of
SMBH growth and galaxy formation, the jets carry an average $\sim 0.4$ of the
SMBH power, while in the non relativistic limit it is a small fraction.

There is a lower limit set by the requirement that the jets have enough energy
to expel the required amount of gas from the newly formed galaxy; the
requirement is that $E_{k,\rm f}\ga\frac{1}{2}M_{\rm G}\sigma^2$, or
\begin{eqnarray}
f\ga\frac{M_{\rm G}\sigma^2}{2\eta M_{\rm BH} c^2}.
\end{eqnarray}
Using the same typical numbers as in Section \ref{sec:intro}, we find $f\ga0.002$.
This is an important result: even if the jets are not relativistic and carry
only a small fraction of the energy, then this mechanism can work.
Non-relativistic wide outflows, termed slow massive wide (SMW) outflows, can be
formed by disk winds instead of highly relativistic jets from very close to the
SMBH, or from a narrow relativistic jet that very close to the SMBH turns into a
non-relativistic jet by interacting with the ambient material \citep{Soker2008}.

Let us elaborate on SMW outflows. Feedback based on slow ($\sim 10^4 \kms$)
massive wide jets has been applied in cooling flow clusters
(\citealt{Sternberg_etal2007}; \citealt{Soker_etal2010}) and in elliptical
galaxies \citep{Ostriker_etal2010} to heat and expel gas.
The properties of such SMW outflows have been recently deduced
(\citealt{Moe_etal2009}; \citealt{Dunn_etal2010}).
For these outflows to be energetic, they must result from accretion close to the
SMBH, rather than from winds, which originate in larger distances (where
the potential well is shallow).
\citet{Soker2008} argued that fast jets launched from the vicinity of the SMBH can form
energetic enough SMW outflows, and that
the condition for the fast jets to power the SMW outflow is that the fast jets
do not penetrate through the gas (as in the penetrating jet mechanism). Namely,
the mechanism studied in this paper to establish the $M_{\rm BH}$--$\mu$
correlation can also explain the energy of SMW outflow. Note that we do not
confront the question of energy transfer from the accretion process to the SMW
outflow.

It is noted again that in the full relation between $M_{\rm BH}$ and $M_{\rm
G}$, the ratio of the transverse velocity to the dispersion would appear squared
(cf. equation \ref{eq:Ms-limit}). It is assumed in the model that $v_{\rm rel} =
\sigma$. However, they need not be exactly equal, but we do expect them to be
comparable. This term will surely introduce a large scatter in the relation; it
might also introduce a systematic shift. In such a case the average value we
derive here for $\eta_p$ incorporates this systematic shift.

\section[Summary]{\caps{Summary}}
\label{sec:summary}

We examined the correlation between $M_{\rm BH}$, and the momentum parameter,
$\mu\equiv M_{\rm G} \sigma/c$. The motivation for this study is the
penetrating-jet feedback mechanism that predicts such a correlation
(Section \ref{sec:model}; \citealt{Soker2009}), although the SMBH determines the
correlation by depositing energy into the ISM rather than momentum. Using the
sample of \citet{Gultekin_etal2009} we examined the correlation with $\mu$ in
Section \ref{sec:correlation}. We applied our results to the penetrating-jet feedback
mechanism in Section \ref{sec:feedback}. The main results of our study and the implied
insights can be summarized as follows:
\begin{enumerate}
\item Despite the large statistical uncertainties, recent data suggest that the
masses of SMBHs are indeed correlated with the momentum parameter of the
bulge. Moreover, we find that the relation is a linear one.
\item The penetrating-jet feedback mechanism \citep{Soker2009} is compatible
with such a relation.
\item In the jet-penetrating feedback mechanism, $\eta_p$ is a fundamental
parameter that must be obtained from observations. It has the same role as
$\eta$ in equation (\ref{eq:eta}), the relation for the total energy released by
the accreting BH, but for the momenta of the two jets. We find $\eta_p \sim
0.04$, which is of course an average value over many systems, and average over
time in each system (during the phase of SMBH growth).
\item This allows us to scale equation (\ref{eq:cor1}) by
\begin{eqnarray}
\frac {M_{\rm G}} {M_{\rm BH}}  = 480
\left( \frac{\sigma}{200 \kms} \right)^{-1}
\left( \frac{\eta_p}{0.04} \right)
\label{eq:corf}
\end{eqnarray}
\item We further analysed the implication of $\eta_p \sim 0.04$, and derived an
expression for the fraction of the SMBH power that is carried by the jets
(equation \ref{eq:f}). While relativistic jets carry $\sim 38$ per cent of the released
power, non relativistic jets carry a small fraction of the power (the rest is in
radiation). Although the latter only carry a small fraction of the total power,
they are still able to maintain the feedback process. In this model, there is no
need for ``fine tuning'' in using only a small fraction of the jets' energy, but
instead a substantial fraction of the energy is used.
\item We cannot determine with current data which of the correlations of $M_{\rm
BH}$ (with $\mu$, $\sigma$, $M_{\rm G}$, or $M_{\rm G}\sigma^2$) is better.
\end{enumerate}

\section*{\caps{Acknowledgements}}\addcontentsline{toc}{section}{Acknowledgements}
We thank Adi Nusser, Ari Laor, and Yoram Rozen for very helpful insights,
and an anonymous referee for comments that improved the manuscript.
This research was supported by the Asher Fund for Space Research at the Technion, and
the Israel Science foundation.

\appendix

\section*{\caps{Appendix}}\addcontentsline{toc}{section}{Appenix}

In the fitting method we used in this paper, the following expression was
minimized to find $\alpha$ and $\beta$: \begin{eqnarray}
\chi^2 = \sum_{i=1}^N \frac{(y_i - \alpha - \beta
x_i)^2}{\epsilon_{yi}^2+\beta^2\epsilon_{xi}^2
+ {\rm E}^2(\beta, \epsilon_0)},
\end{eqnarray}
where $x_i$ and $y_i$ are the data points, $\epsilon_{xi}$ and $\epsilon_{yi}$
are the uncertainty values for galaxy $i$; ${\rm E}^2(\beta, \epsilon_0)$ is
different in the three scatter model used
\begin{eqnarray}
{\rm E}^2(\beta, \epsilon_0) = \begin{cases}
\beta^2\epsilon_0^2 & x\textrm{-scatter}\\
\epsilon_0^2 & y\textrm{-scatter}\\
\epsilon_0^2(1+\beta^4)/(1+\beta^2) & \textrm{orthogonal scatter.}\end{cases}
\end{eqnarray}
If no intrinsic scatter is considered, then ${\rm E}^2=0$ and the $\chi^2$
estimator is symmetric in $x$ and $y$, as noted by \citet{Tremaine_etal2002}.
When adding the intrinsic scatter, only the orthogonal model preserves this
feature. In the results shown in Fig. \ref{fig:slope}, we
increase (from zero) the value of
$\epsilon_0$ so that $\chi^2$ per degree of freedom (denoted by $\chi_r^2$)
approaches its expectation value of unity. When demanding $\beta=1$ (as we do
in obtaining equation \ref{eq:alpha}) the three models become identical. The
calculation of the formal error on
the best fitting parameters is based on finding an $(\alpha,\beta)$ pair for
which $\chi^2$ is larger by 1 from its minimal value, as detailed in great
length in \citet{Press_etal1992}.

One can think of orthogonally scattered data in the following way: a certain law
of nature produces a linear relation between $x$ and $y$, but various other
natural processes cause objects to slightly move on the $xy$ plane. Points
$(x_i,y_i)$ with no measurement errors will have distances $d_i$ from the ridge
line of the relation, which are distributed normally with variance
$\epsilon_0^2$. Note that while the units of $\epsilon_0$ are understandable in
the other models (as they correspond to one axis), here the units of the
intrinsic scatter are intermediate between the two coordinates. Thus, the
obtained value of $\epsilon_0$ is only meaningful in a particular plane, and
cannot be used to compare goodness of fit of different relations.

We did not study the orthogonal estimator thoroughly, and it is proposed here
just as a logical compromise between the two other choices. we briefly describe
the behaviour of this estimator: when the slope is high, the orthogonal scatter
model behaves like the $x$-scatter model, and in fact one usually gets a
slightly higher best fitting $\beta$; when the slope is too low, then the
estimated $\beta$ is close to that obtain by the $y$-scatter model. Therefore,
it seems that this approach is good primarily when the slope is moderate, which
was the case studied here.

\label{lastpage}

\end{document}